\documentstyle[11pt,newpasp,epsf,twoside]{article}
\begin{document}
\title{The GEO\,600 Gravitational Wave Detector -- Pulsar Prospects}
 \author{G.~Woan$^{1}$,
P.~Aufmuth$^{2}$, C.~Aulbert$^{4}$, S.~Babak$^{5}$,
R.~Balasubramanian$^{5}$,  B.\,W.~Barr$^{1}$, S.~Berukoff$^{4}$,
S.~Bose$^{4}$, G.~Cagnoli$^{1}$, M.\,M.~Casey$^{1}$,
D.~Churches$^{5}$, C.~N. Colacino$^{2}$, D.\,R.\,M.~Crooks$^{1}$,
C.~Cutler$^{4}$, K.~Danzmann$^{2,\,3}$, R.~Davies$^{5}$,
R.\,J.~Dupuis$^{1}$, E.~Elliffe$^{1}$, C.~Fallnich$^{6}$,
A.~Freise$^{3}$, S.~Go{\ss}ler$^{2}$, A.~Grant$^{1}$,
H.~Grote$^{3}$, G.~Heinzel$^{2}$, A.~Hepstonstall$^{1}$,
M.~Heurs$^{2}$, M.~Hewitson$^{1}$, J.~Hough$^{1}$,
O.~Jennrich$^{1}$, K.~Kawabe$^{3}$, K.~K\"otter$^{2}$,
V.~Leonhardt$^{2}$, H.~L\"uck$^{2,\,3}$, M.~Malec$^{2}$,
P.\,W.~McNamara$^{1}$, K.~Mossavi$^{3}$, S.~Mohanty$^{4}$,
S.~Mukherjee$^{4}$, S.~Nagano$^{2}$, G.\,P.~Newton$^{1}$,
B.\,J.~Owen$^{4}$, M.\,A.~Papa$^{4}$, M.\,V.~Plissi$^{1,\,4}$,
V.~Quetschke$^{2}$, D.\,I.~Robertson$^{1}$,
N.\,A.~Robertson$^{1}$, S.~Rowan$^{1}$, A.~R\"udiger$^{3}$,
B.\,S.~Sathyaprakash$^{5}$, R.~Schilling$^{3}$,
B.\,F.~Schutz$^{4,\,5}$, R.~Senior$^{5}$, A.\,M.~Sintes$^{8}$,
K.\,D.~Skeldon$^{1}$, P.~Sneddon$^{1}$, F.~Stief$^{2}$,
K.\,A.~Strain$^{1}$, I.~Taylor$^{5}$, C.\,I.~Torrie$^{1}$,
A.~Vecchio$^{4,7}$, H.~Ward$^{1}$, U.~Weiland$^{2}$,
H.~Welling$^{6}$, P.~Williams$^{4}$, W.~Winkler$^{3}$,
B.~Willke$^{2,\,3}$, I.~Zawischa$^{6}$}
 \affil{
$^{1}$\,%
  Department of Physics \& Astronomy, University of Glasgow,
  Glasgow,~G12~8QQ, UK\\[0mm]
$^{2}$\,%
  Institut f\"ur Atom- und Molek\"ulphysik,
  Universit\"at Hannover, Callinstr.~38,
  30167~Hannover,~Germany\\[0mm]
$^{3}$\,%
  Max-Planck-Institut f\"ur Gravitationsphysik,
  Albert-Einstein-Institut Hannover,
  Callinstr.~38,~30167~Hannover,~Germany\\[0mm]
$^{4}$\,%
  Max-Planck-Institut f\"ur Gravitationsphysik,
  Albert-Einstein-Institut Golm,
  Am~M\"uhlenberg~1,
  14476~Golm,~Germany\\[0mm]
$^{5}$\,%
  Department of Physics and Astronomy,
  Cardiff University,
  P.O.~Box~913,
  Cardiff, CF2~3YB, UK\\[0mm]
$^{6}$\,%
  Laser Zentrum Hannover~e.\,V.,
  Hollerithallee~8,
  30419~Hannover,~Germany\\[0mm]
$^{7}$\,%
  School of Physics and Astronomy,
  University~of~Birmingham,
  Edgbaston, Birmingham,~B15~2TT,~UK\\[0mm]
  $^{8}$\,%
  Departament~de~Fisica,
  Universitat~de~les~Illes~Balears,
  E-07071~Palma~de~Mallorca,~Spain }

\begin{abstract}
The GEO\,600 laser-interferometric gravitational wave detector
near Hannover, Germany, is one of six such interferometers now
close to operation worldwide.  The UK/German GEO collaboration
uses advanced technologies, including monolithic silica
suspensions and signal recycling, to deliver a sensitivity
comparable with much larger detectors in their initial
configurations.  Here we review the design and performance of
GEO\,600 and consider the prospects for a direct detection of
continuous gravitational waves from spinning neutron stars.
\end{abstract}
\section{Introduction}
The world-wide effort to detect gravitational waves  directly has
resulted in the construction of a network of laser
interferometers. The US LIGO project (Sigg 2002) contributes three
instruments, and the French-Italian VIRGO detector (Acernese et
al.\ 2002), the TAMA\,300 detector in Japan (Kuroda et al.\ 2002)
and the UK/German GEO\,600 detector (Willke et al.\ 2002) one
each. These detectors are based on the principles of Michelson
interferometry and use two long orthogonal arms to form a
quadrupole antenna sensitive to gravitational waves, although the
design details vary considerably between detectors.

The GEO\,600 detector at Ruthe (close to Hannover, Germany) is now
nearly complete and has recently carried out its first science
observations.  These were performed in coincidence with the LIGO
detectors as part of its role within the LIGO Science
Collaboration (LSC).

\section{Interferometer design and performance}
GEO was constructed on a site that restricted its arm lengths to
600\,m, so considerable effort was put into its mechanical and
optical design to reach the desired sensitivity.  Seismic
disturbances and thermal motions of the mirrors and of other
optical components introduce changes in the optical path lengths
along the two arms and add noise to the differential strain, $h$,
caused by a passing gravitational wave. To minimise this the
entire optical path, including the 60\,cm diameter tube arms, is
held at high vacuum ($\sim10^{-9}$\,mbar), and materials capable
of significant outgassing are sealed in steel or glass.

The calibrated strain output of the detector has many contributing
noise components, each with a characteristic spectral density.
Below about 40 Hz the spectrum is dominated by seismic noise, and
seismic isolation is provided by passive stainless steel and
rubber stacks supporting double or triple pendulum suspensions
holding the optical components (Plissi et al.\ 2000). The beam
splitter, end mirror optics and lower suspensions are made of
fused silica, suspended by 270\,$\mu$m diameter fused silica
fibres, joined with welds and hydroxy-catalysed bonds (Rowan et
al.\ 1998). These high-strength bonds preserve the homogeneity of
the silica and give the lower assemblies mechanical quality
factors in excess of $10^7$ (Cagnoli et al.\ 2000).  At
frequencies above about 400\,Hz the dominant spectral noise power
is from laser shot noise, whilst at intermediate frequencies the
spectrum is dominated by the thermal and thermo-refractive noise
intrinsic to silica (Braginsky, Gorodetsky \& Vyatchanin 2000).

Further noise can be introduced by laser fluctuations or changes
in optical alignment causing modulation of the light power at the
interferometer's output port. The laser system is built around a
master non-planar ring oscillator that is injection-locked to a
diode-pumped Nd:YAG slave laser delivering 12\,W at 1064\,nm.  The
light is fed to two sequential mode-cleaners that temporally and
spatially filter the beam and seismically isolate it. The
Michelson interferometer is held on a dark fringe at its output
port, and configured as a resonant cavity by inserting a power
recycling mirror at the input port. This builds up the effective
power at the beam splitter to several kilowatts and so improves
the shot-noise limited sensitivity of the instrument.

Although the detector is intrinsically broadband we can trade
bandwidth for sensitivity by `signal recycling' (Meers  1988).
When installed, this will increase the shot-noise limited
sensitivity by a factor of about 10 around a chosen frequency and
promises to be of particular importance for targeted observations
of low-mass X-ray binaries (Bildsten, these proceedings).

\section{Observing programme}
The sensitivity of GEO  (and of all the current generation of
detectors) is such that the probability of an astrophysical
detection is fair.  We will have to wait until Advanced LIGO
before the odds become well stacked in our favour.  However the
prospect of a detection is very real, and specific effort has gone
into optimal strategies for detecting the chirp waveforms from
black-hole/black-hole inspirals, burst signals coincident at
several detectors, the stochastic gravitational wave background
and signals from continuous wave sources such as pulsars.

\begin{figure}
 \plottwo{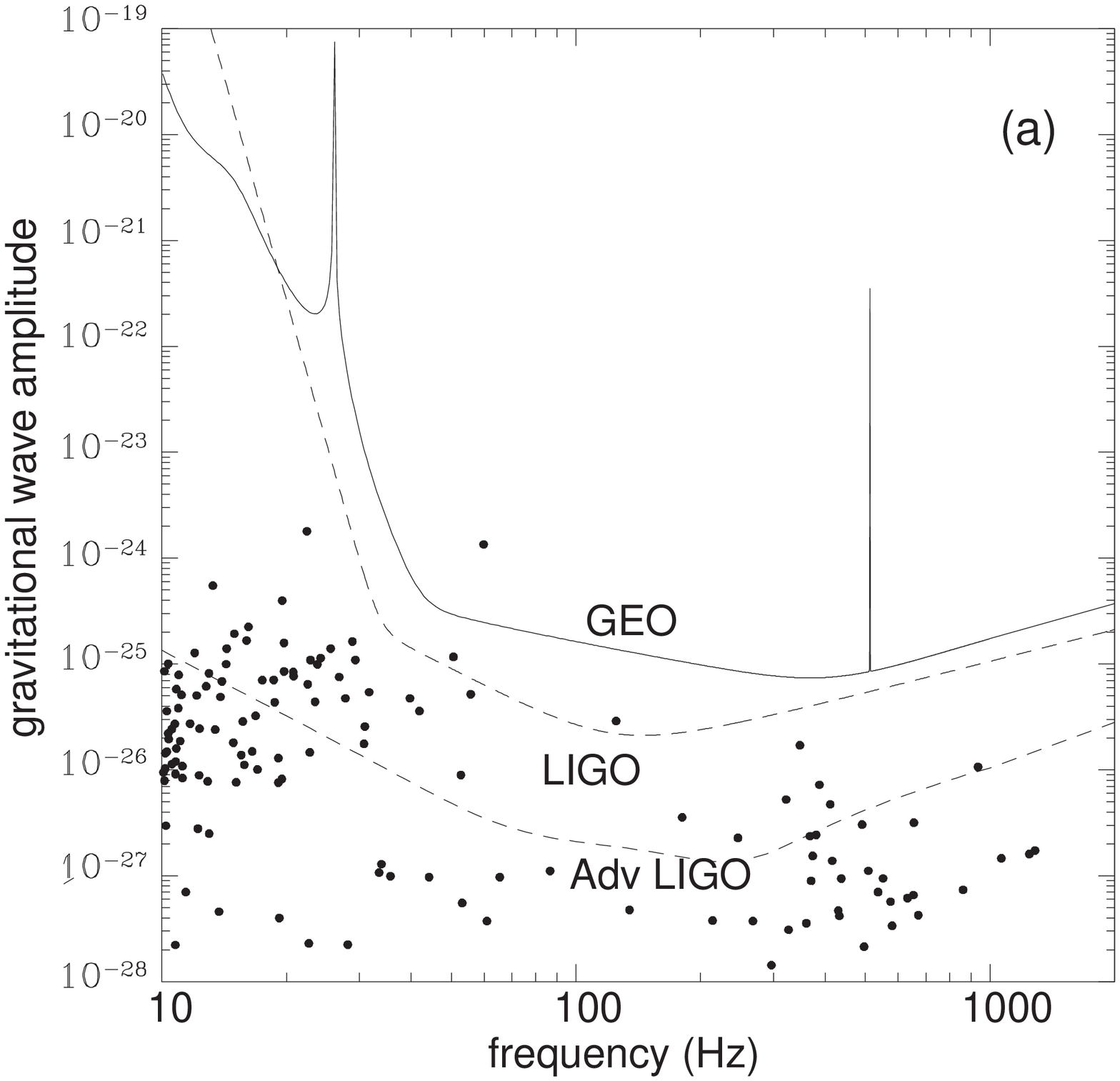}{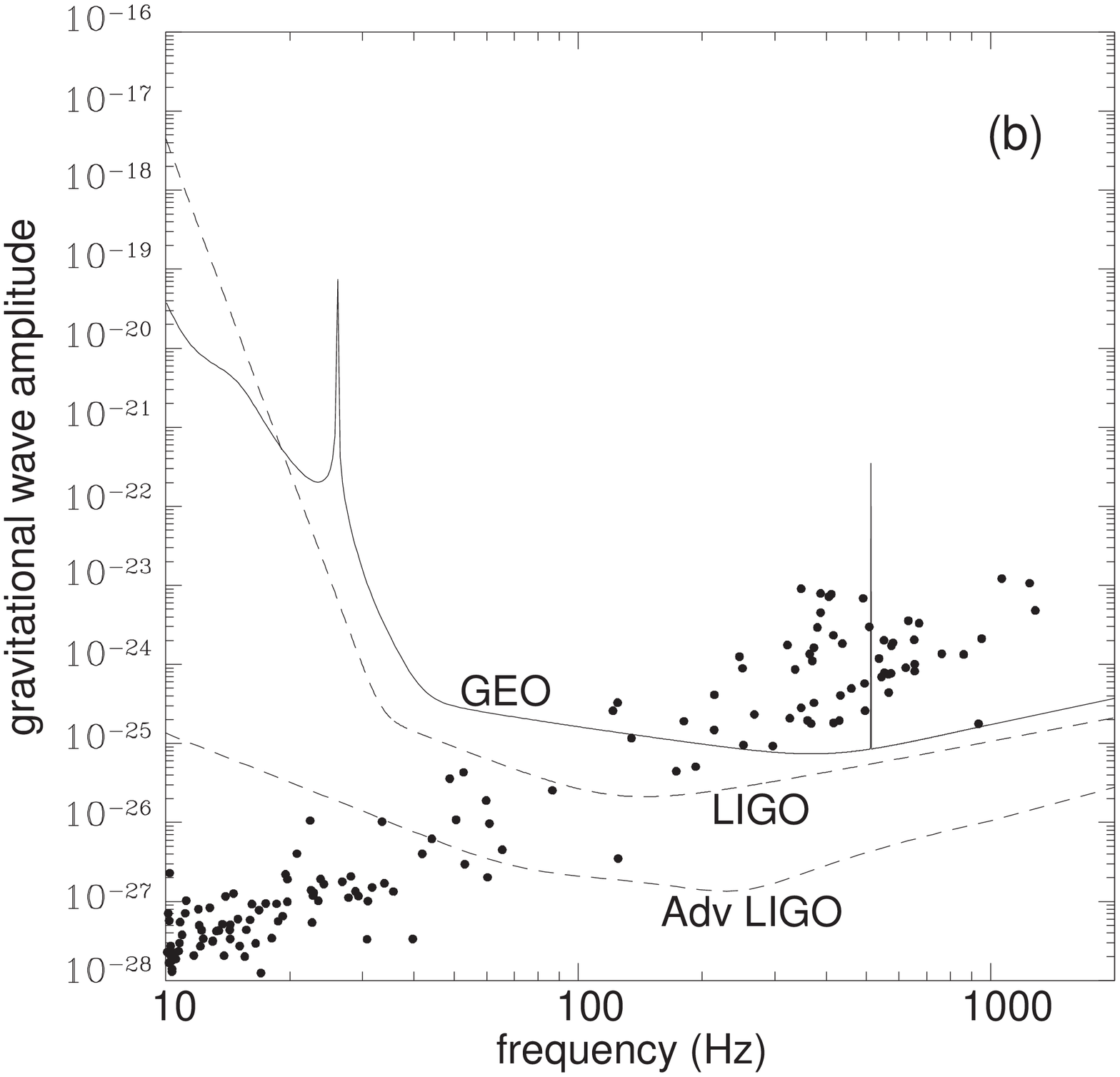}
 \caption{Theoretical strain signals from known pulsars after one
year of integration, (a) attributing the rate of loss of
rotational kinetic energy to gravitational luminosity and (b)
assuming an equatorial ellipticity of $10^{-5}$. The one-year
theoretical sensitivities for GEO, LIGO and Advanced LIGO are also
shown. }
\end{figure}

We expect a non-precessing neutron star of equatorial ellipticity
$\epsilon$ spinning at a frequency $\nu_0$ to produce a strain
amplitude, $h_0$, a distance $r$ away of
\begin{equation}
h_0 = \frac{16\pi^2 G}{c^4}\frac{I_{zz}\nu_0^2}{r}\epsilon,
\end{equation}
where $I_{zz}$ is the principal moment of inertia about the
rotation axis, $G$ is the gravitational constant and $c$ the speed
of light (e.g., Brady 1998). It is interesting to note that
because our detection is coherent, the signal strength falls off
with distance only as $1/r$. The interferometer has antenna
response patterns, $F_+$ and $F_\times$, to the `$+$' and
`$\times$' wave polarisations. For a given source, these antenna
responses depend both on time, $t$, (this is a transit instrument)
and on the polarisation angle of the source ($\psi$, loosely its
position angle around the line of sight), so the detected strain
signal is quasi-sinusoidal and of the form
\begin{equation}
h(t)=\frac{1}{2}F_+(\psi,t)h_0(1+\cos^2\iota)\cos2\Phi(t) +
F_{\times}(\psi,t)h_0\cos\iota\sin2\Phi(t),
\end{equation}
where $\iota$ is the angle between the spin axis and the line of
sight, and $\Phi(t)$ is the observed rotational phase
($\simeq2\pi\nu_0t$) (e.g., Jaranowski et al.\ 1998).  The
detector has a quadrupole response to each polarisation, and the
Earth is transparent to gravitational waves, so nearly the whole
celestial sphere is under observation at any one time.  The
largest uncertainties in the expected strain signal are the
neutron star's equatorial ellipticity and, for radio-quiet neutron
stars, their distances and frequencies.  We can see from Figure~1
and Equation~1 that millisecond pulsars represent  some of our
best candidates for a direct detection. Predicted sensitivities
require a millisecond source at 10\,kpc to have an ellipticity of
$\sim10^{-6}$ (a quadrupole moment of $\sim 10^{39}$\,g\,cm$^2$)
for detection after one year. However the gravitational luminosity
would need to come from a source somewhat better than the observed
rate of decrease of rotational kinetic energy for these objects
(Figure 1a), if we take them to be isolated and effectively rigid
rotators.

\section{The future}
Both GEO and LIGO are presently on their final commissioning
phases, and are expected to be working at full sensitivity by
2003. Members of GEO and LIGO are already working within the LSC
to share software and data. Present pulsar-related efforts in GEO
are concentrating on targeted searches for emission from known
radio pulsars, although the algorithmic and computational
resources needed to carry out more general all-sky searches are
well advanced.

\acknowledgments The authors would like to thank the Particle
Physics and Astronomy Research Council in the UK (PPARC), the
German Bundesministerium f\"ur Bildung und Forschung (BMBF) and
the State of Lower Saxony (Germany) for their funding and
continuous support of the GEO\,600 project.

\end{document}